\documentclass[12pt]{iopart}

\usepackage{ulem}
\usepackage{color}
\usepackage{graphicx}
\usepackage{epsfig}
\usepackage{bm}

\def\nn {\nonumber}

\newcommand{\be}{\begin{equation}}
\newcommand{\ee}{\end{equation}}
\newcommand{\bea}{\begin{eqnarray}}
\newcommand{\eea}{\end{eqnarray}}

\newcommand{\om}{\omega}

\newcommand{\ks}{k \!\!\! /}

\newcommand{\ls}{l \!\!\! /}

\newcommand{\vk}{\vec k}

\newcommand{\vq}{\vec q}

\newcommand{\mn}{\mu\nu}

\begin{document}

\title[]{Calculations of Shear, Bulk viscosities and Electrical conductivity in the Polyakov-Quark-Meson 
model}

\author{Pracheta Singha$^{1,a}$, Aman Abhishek$^{2,3b}$, Guruprasad Kadam$^{4,c}$, 
Sabyasachi Ghosh$^{5,d}$, Hiranmaya Mishra$^{2,e}$}

\address{$^1$Center for Astroparticle Physics and Space Science, Bose Institute, 
Block-EN, Sector-V, Salt Lake, Bidhan Nagar, Kolkata - 700091, India}
\address{$^2$Theory Division, Physical Research Laboratory, 
Navrangpura, Ahmedabad 380 009, India}
\address{$^3$Indian Institute of Technology Gandhinagar, Palaj, 
Gandhinagar 382355, Gujarat, India}
\address{$^4$Department of Physics, Shivaji University, Kolhapur, 
Maharashtra-416004, India}
\address{$^5$Indian Institute of Technology Bhilai, GEC Campus, Sejbahar, Raipur-492015, 
Chhattisgarh, India}

\ead{$^a$pracheta.singha@gmail.com, $^b$aman@prl.res.in, 
$^c$guruprasadkadam18@gmail.com, $^d$sabyaphy@gmail.com, 
$e$hm@prl.res.in}

\vspace{10pt}

\begin{abstract}
We have calculated different transport coefficients like shear, bulk
viscosities and electrical conductivity of quark
and hadronic matter within the framework of 2 flavor Polyakov-Quark-Meson model.
For constant thermal widths of quarks and mesons, the temperature
dependence of different transport coefficients reveals the thermodynamical
phase space structure and their qualitative behavior are quite well in agreement
with earlier works, based on other dynamical models. Besides the
phase-space structure of quarks and mesons, their thermal width also
have explicit temperature dependence. This dependence has been obtained here
from the imaginary part of their respective
self-energies at finite temperature. Due to the threshold conditions
of their self energies, only some limited temperature regions of quark and hadronic phase
are relevant for our numerical estimation
of transport coefficients, which are grossly in agreement with some of the earlier results.
An interesting outcome of the present analysis is that the quark relaxation time due to quark-meson loops can develop
perfect fluid nature within a specific temperature range.
\end{abstract}

%
%
%
%
%

\section{Introduction}
Microscopic calculations of transport coefficients
like the shear and bulk viscosities of quark and hadronic
matter are one of the contemporary research interests in
the field of heavy ion physics.
 These coefficients are relevant not only because
they enter as inputs for dissipative hydrodynamical simulations,
but also, through their dependence on system parameters like
temperature and chemical potential, they can indicate the location of phase transition
in the phase diagram
~\cite{Kapusta:2008vb}. Indeed, the small value of the shear viscosity to entropy density ratio $\eta/s$ to
explain the data of elliptic flow~\cite{romatschke,gale} and its connection with the Kovtun-Son-Starinets (KSS)
lower bound, $\eta/s=1/4\pi$,  has spurred many activities to investigate this coefficient
theoretically~\cite{{Kapusta:2008vb},{Arnold},{Nills},{Meyer_eta},{eta_LQCD},{Itakura},{Dobado},{Nicola},{Weise1},{SSS},{Ghosh_piN},{Gorenstein},{HM},{HM2},{Hostler},{Purnendu},{Redlich_NPA},{Marty},{G_IFT},{Deb},{Tawfik},{G_CAPSS},{Weise2},{Sedarkian},{HM_PQM2},{Kinkar},{Bass},{Muronga},{Plumari},{Pal},{Sadooghi},{Albright},{Prakash},{Gavin}}.
Among them, Refs.~\cite{Arnold,Nills,Meyer_eta} and
Refs.~\cite{Itakura,Dobado,Nicola,Weise1,SSS,Ghosh_piN,Gorenstein,HM,HM2,Hostler}
have calculated $\eta$ of
quark matter and hadronic matter respectively, whereas
Refs.~\cite{{Purnendu},{Redlich_NPA},{Marty},{G_IFT},{Deb},{Tawfik},{G_CAPSS},{Weise2},{Weise3},{Sedarkian},{HM_PQM2},{Kinkar}}
have calculated $\eta$ for both the phases
covering the entire range of temperature ($T$). Some simulation based
calculations for $\eta$ are addressed in Refs.~\cite{Bass,Muronga,Plumari,Pal}.
These investigations provide a grossly settled picture of the temperature dependence
of viscosity to entropy density ratio ($\eta/s$) which has a minimum near
transition temperature $T_c$,
similar to helium, nitrogen, and water~\cite{Kapusta:2008vb}.
However the numerical results of these calculations near
critical temperature seem to differ by order of
magnitude. For example Refs.~\cite{Itakura,Weise1},
Refs.~\cite{Nicola,Prakash,SSS} and Refs.~\cite{Dobado} have
predicted $\eta \approx 0.001$~GeV$^3$, $\eta \approx 0.002-0.003$~GeV$^3$ and $\eta \approx 0.4$~GeV$^3$ respectively.

Similar to the shear viscosity,  bulk viscosity ($\zeta$) has a long list of references
as well~\cite{{Nicola},{HM},{HM2},{Hostler},{Purnendu},{Redlich_NPA},{Marty},{G_IFT},{Deb},{Tawfik},{Prakash},{Gavin},{Paech1},{Arnold_bulk},{Santosh},{Meyer_zeta},{Dobado_zeta1},{Dobado_zeta2},{Redlich_PRC},{De-Fu},{Tuchin},{Tuchin2},{Vinod},{Nicola_PRL},{Sarkar},{SG_NISER},{Sarwar},{Kadam10},{Kinkar_bulk}},
where most of the calculations are done for zero baryonic chemical potential.
Hard thermal loop (HTL) calculations of $\zeta$ have been done by  Ref.~\cite{Arnold_bulk}.
The calculations of effective models of quantum chromodynamics (QCD) like
Nambu-Jona-Lasinio (NJL) model~\cite{Redlich_NPA,Marty,G_IFT,Deb,De-Fu,Kinkar_bulk}
and linear sigma model (LSM)~\cite{Purnendu,Paech1,Dobado_zeta2}
have explored the temperature dependence of $\zeta$ for both quark and hadronic phases.
Refs.~\cite{Nicola,HM,HM2,Hostler,Nicola_PRL,Sarkar,Sarwar} are the effective hadronic-model
calculations valid for the hadronic phase only. Some investigations 
both in QGP phase~\cite{{Meyer_zeta},{Tuchin},{Tuchin2}}
and in hadronic phase~\cite{{HM},{Hostler}}
have observed an increasing nature 
of $\zeta/s $ near $T_c$,  which might be associated with
the maximum violation of conformal symmetry at the transition point in the full temperature ~\cite{{Purnendu},{Dobado_zeta2},{Kinkar_bulk}}.
However, instead of this peak structure, Refs.~\cite{{HM2},{Deb},{SG_NISER}}
have observed decreasing nature of $\zeta/s(T)$ with temperature.
Further, the order of magnitude of  $\zeta$ and $\zeta/s$
from different model calculations varies widely
covering a range of values from $ 10^{-5}$ GeV$^3$~\cite{Sarkar}
to $10^{-2}$ GeV$^3$~\cite{Redlich_NPA}
and $10^{-3}$~\cite{Sarkar} to $10^{0}$~\cite{Redlich_NPA} respectively.
Thus it is observed that there are uncertainties in behavior as well
as in  numerical estimations of $\zeta(T)$ and  $\zeta/s(T)$, which require further
investigation.

Another transport coefficient, we are interested in, is the electrical conductivity $\sigma$,
which has been studied by a large number of Lattice QCD
calculations~\cite{LQCD_Ding,LQCD_Arts_2015,LQCD_Arts_2007,LQCD_Buividovich,LQCD_Burnier,LQCD_Gupta,LQCD_Barndt,LQCD_Amato},
predicting a wide band in their numerical results.
Besides these first principle based calculations, some simulation based
calculations using transport codes ~\cite{Cassing,Puglisi,Greif} and other model dependent
calculations~\cite{{Marty},{PKS},{Finazzo},{Lee},{Nicola_PRD},{Greif2},{G_IFT2},{VC},{G_CU}}
for $\sigma$ have also been done.
Most of the earlier
works~\cite{Marty,Cassing,Puglisi,Greif,PKS,Finazzo,Nicola_PRD,Greif2,G_CU}
have observed decreasing nature of the ratio $\sigma(T)/T$ in hadronic phase~\cite{Marty,Cassing,Nicola_PRD,Greif2,G_CU}
and increasing behavior with temperature in the QGP phase~\cite{Marty,Cassing,Puglisi,Greif,PKS,Finazzo,LQCD_Amato}.
On the other hand, some studies e.g.  Refs.~\cite{Finazzo,Lee,G_IFT2} have shown that
$\sigma/T$ increases with $T$ in the hadronic phase.
Not only the general temperature dependence of this dimensionless ratio $(\sigma/T)$, the uncertainty
also appears in the numerical values,
whose approximate range may be considered as $\sigma/T\approx 10^{-3}$ to $10^{-2}$
for hadronic phase and $\sigma/T\approx 10^{-3}$ to $10^{-1}$ for quark phase.

Above discussions point to the fact that the numerical values
of the transport coefficients, obtained from different microscopic calculations,
vary in a large band. In this context, $\eta/s$ may be a good guiding candidate
to provide a possible converging picture because RHIC and LHC experiments indicate that $\eta/s$ of the medium should be very close
to the KSS bound. In this regard, quark-meson interactions
near transition temperature may be physically a better picture for creating a near
perfect fluid because as the coupling is quite strong. Refs.~\cite{Redlich_NPA,Marty,G_IFT,Deb,G_CAPSS,Weise2,Weise3,Sedarkian}
in NJL model, Ref.~\cite{HM_PQM2} in PQM model have concluded that quark-meson
interaction is very important near transition temperature for building up
the perfect fluid nature of the medium.
In the present article, we have estimated different
transport coefficients in the framework of 2 flavor Polyakov-Quark-Meson (PQM) model, where
we have evaluated one-loop self-energy diagrams of quark and mesons to
obtain their relaxation times in terms of thermal widths. Unlike 
NJL model~\cite{Redlich_NPA,Marty,G_IFT,Deb,G_CAPSS,Weise2,Weise3,Sedarkian},
here in PQM model, we have the explicit meson degrees of freedom which can also contribute to the transport
coefficients along with quark component. With respect to earlier work on PQM model by some of the authors of present investigation~\cite{HM_PQM2}
, the main difference here lies in calculations of quark (Q) and
meson (M) relaxation times. Ref.~\cite{HM_PQM2} basically considered $2\rightarrow2$
type of scattering processes like $QQ\rightarrow QQ$, $QM\rightarrow QM$,
$MM\rightarrow MM$. In the present work, on the other hand, we calculate $1\rightarrow 2$
in-elastic kind of diagrams to obtain the relaxation times of quarks and mesons.
Eventually, we will observe, for some particular temperature range,
such in-elastic scatterings play an important role in dissipation.
The forward $Q\rightarrow QM$ and reverse $QM\rightarrow Q$ scatterings can
be obtained from in-medium quark self-energy diagram, having different $QM$ loops. In hadronic phase, $\pi$ and $\sigma$ meson
self-energies with $\pi\sigma$ and $\pi\pi$ loops can give
$(\pi/\sigma)\leftrightarrow \pi(\sigma/\pi)$ kind of forward and reverse scatterings.
We will find that these in-elastic scatterings, in certain regions of temperature,
can help to build perfect fluid nature in the medium.

The paper is organized as follows. The next section
addresses the formalism part of the PQM model and transport coefficients. The analytic structure of the
self-energies and their contributions to the transport coefficients of quark and hadronic matter are
rigorously discussed in the result section, followed by the summary
and conclusions in the last section.

\section{Formalism}

\subsection{Thermodynamics of two-flavor PQM model and  meson masses}
To incorporate aspects of chiral symmetry breaking and its restoration in
a medium as well as confinement-deconfinement transition,
we shall adopt here the Polyakov loop extended quark meson model.
This is an extension of the linear sigma model that provides
an effective realization of chiral symmetry. Coupling the quarks and meson degrees
of freedom to the expectation values of the Polyakov loop,
the physics of confinement is expected to be taken into account here. We confine the
investigation here regarding the transport coefficients to the two flavor version of
the PQM model. The corresponding Lagrangian density is given as
\bea
&&{\cal L}=\bar\psi\left(i\gamma^\mu D_\mu -g(\sigma+i\gamma_5\vec\tau\cdot\pi)\right)\psi
+\frac{1}{2}\left[\partial_\mu\sigma\partial^\mu\sigma+\partial_\mu\vec\pi\partial^\mu\vec\pi\right]
\nn\\
&&~~~~~~~~~~~~~~~
-U_\chi(\sigma,\vec\pi)-U_P(\Phi,\bar\Phi)~.
\label{lagpqm}
\eea
Here  $\psi=(u,d)$ is a SU(2)$_f$ isodoublet interacting with the isovector ($\sigma$, $\vec\pi$) field.
 The quark field is also coupled to a spatially constant temporal gauge field $A_0$
through the covariant derivative $D_\mu=\partial_\mu-ieA_\mu$; $A_\mu=\delta_{\mu 0}A_\mu$.
The mesonic potential $U_\chi(\sigma,\vec \pi)$ essentially describes the chiral symmetry breaking and is given by
\be
U_\chi(\sigma,\vec \pi)=\frac{\lambda}{4}(\sigma^2+\vec\pi^2-v^2)^2-C\sigma~.
\label{uchi}
\ee
The parameters of the mesonic potential  are chosen so that the chiral symmetry is spontaneously broken in the vacuum
and the expectation values of the meson fields are $\langle\sigma\rangle=f_\pi$, $\langle\vec \pi\rangle=0$,
where $f_\pi=93$ MeV is the pion decay constant. The constant $C$ is fixed from partial conserved
axial current leading to $C=f_\pi m_\pi^2$, with $m_\pi=138$ MeV
being the pion mass.
$v^2=f_\pi^2-m_\pi^2/\lambda$ is obtained by minimizing the potential. The coupling $\lambda$ is fixed
from the sigma mass $m_\sigma^2=m_\pi^2+2\lambda f_\pi^2$. With $m_\sigma=600 $ MeV leads
to $\lambda=\frac{(m_{\sigma}^2-m_{\pi}^2)}{2f_{\pi}^2}\simeq 19.7$.
The  Yukawa coupling $g$ is fixed from the requirement that the constituent quark mass in vacuum
$M_Q=g f_\pi$. With $M_Q=300$ MeV, one gets  $g\simeq 3.3$.


The Polyakov loop potential $U_P(\Phi,\bar\Phi)$ in the Lagrangian in Eq.(\ref{lagpqm}) includes the physics
of color confinement. The Polyakov loop variable
$\Phi$=$\Phi(\vec{x})$=$\frac{1}{N_c}\langle {\rm tr}_c L (\vec{x}))\rangle_\beta $,
where, the Wilson line $L(\vec{x})$ in the temporal direction is given as, with $\beta$=$\frac{1}{T}$ :

\be
 L=P \exp\left( i g \int_0^\beta dx_0 A_0(x_0,\vec x)\right)
\ee
 where $P$ denotes path ordering and $\tau$ is the imaginary time, $\tau$:0 $\to$ $\beta$. The variable $\Phi$ is an order parameter for
confinement-deconfinement transition in the infinitely  heavy quark limit. It vanishes in the confined phase
and attains a non-zero value in deconfined phase. The explicit form of the
potential $U_P(\Phi,\bar\Phi)$ is not known from first principle calculations and the following
is a fit taken from lattice results \cite{BJ}
\be
U_P(\Phi,\bar\Phi)=T^4\left[-\frac{b_2(T)}{2}\bar\Phi\Phi-\frac{b_3}{2}(\Phi^3+\bar\Phi^3)
+\frac{b_4}{4}(\bar\Phi\Phi)^2\right]
\label{uphi}
\ee
with the  coefficients given as
$b_2(T)=6.75-1.95(\frac{T_0}{T})+2.625(\frac{T_0}{T})^2-7.44(\frac{T_0}{T})^3 ,~b_3=0.75,~ b_4=7.5.$
The parameter $T_0$ corresponds to the transition temperature of Yang-Mills theory. However, for the full
dynamical QCD, there is a flavor dependence on $T_0(N_f)$. For two flavors we take it
to be $T_0(N_f=2)=192$ MeV as in Ref.\cite{BJ}.

To calculate the bulk thermodynamical properties of the system we use
a mean field approximation for the mesons and the Polyakov loop fields while retaining the
quantum and thermal fluctuations of the quark fields. The thermodynamic potential can then be written as
\be
\Omega(T,\mu)=\Omega_{\bar Q Q}+U_\chi+U_P(\Phi,\bar\Phi)~.
\ee
The fermionic part of the thermodynamic potential is given as
\bea
\Omega_{\bar Q Q}&=&-2N_fT\int\frac{d^3p}{(2\pi)^3}
\bigg[\ln\{1+3(\Phi+\bar\Phi e^{-\beta\omega_-})e^{-\beta\omega_-}
+e^{-3\beta\omega_-}\}
\nn\\
&&~~~~~~  +\ln\{1+3(\Phi+\bar\Phi e^{-\beta\omega_+})e^{-\beta\omega_+}
+e^{-3\beta\omega_+}\}\bigg]
\eea
modulo a divergent vacuum part. In the above $\omega_\mp=E_p\mp\mu$, with the single
particle quark/anti-quark energy $E_p=\sqrt{\vec p^2+M_Q^2}$.
Though we describe the formalism part at finite quark chemical potential $\mu$
but we take it as zero when we describe our studies in result section.
The mean fields are obtained by minimizing $\Omega$ with respect to $\sigma$,
$\Phi$ and $\bar\Phi$. That is
$\frac{\partial\Omega}{\partial\sigma}=\frac{\partial\Omega}{\partial\Phi}
=\frac{\partial\Omega}{\partial\bar{\Phi}}=0$

The $\sigma$ and $\pi$ masses are given by the curvature of $\Omega$ at the global minimum by
$M_\sigma^2=\frac{\partial^2\Omega}{\partial\sigma^2}$ and $M_{\pi_i}^2=\frac{\partial^2\Omega}{\partial\pi_i^2}.$
These equations lead to  the masses of the $\sigma$ and $\pi$ given as
\be
M_\sigma^2=m_\pi^2+\lambda(3\sigma^2-f_\pi^2)+g^2\frac{\partial\rho_s}{\partial\sigma}~,
\label{sigmass}
\ee
with
\be
\rho_s=6N_fg~\sigma\int\frac{d^3 p}{(2\pi)^3}\frac{1}{E_P}\left [f^-_\Phi + f^+_\Phi\right]
\label{rhos}
\ee
and
\be
M_\pi^2=m_\pi^2+\lambda(\sigma^2-f_\pi^2)+g_\sigma^2\frac{\partial\rho_{ps}}{\partial\pi}~,
\label{pimass}
\ee
with
\be
\vec\rho_{ps}=\langle\bar q\i\gamma_5\vec\tau q\rangle=6N_fg~\vec\pi
\int\frac{d^3 p}{(2\pi)^3}\frac{1}{E_P}\left [f^-_\Phi+f^+_\Phi\right].
\label{rhops}
\ee

In the above, $f^\mp_\Phi$ are the distribution functions for the quarks and anti quarks, given as
\bea
f^-_\Phi&=&\frac{\Phi e^{-\beta\omega_-}+2\bar\Phi e^{-2\beta\omega_-}
+ e^{-3\beta\omega_-}}{1+3\Phi e^{-\beta\omega_-}+3\bar\Phi e^{-2\beta\omega_-} + e^{-3\beta\omega_-}}~,
\nn\\
&&{\rm and}
\nn\\
f^+_\Phi&=&\frac{\bar\Phi e^{-\beta\omega_+}+2\Phi e^{-2\beta\omega_+}
+ e^{-3\beta\omega_-}}{1+3\bar\Phi e^{-\beta\omega_+}+3\Phi e^{-2\beta\omega_+} + e^{-3\beta\omega_+}}~.
\label{P_FD}
\eea
We have set the expectation value of pion field to be zero, i.e. $\vec\pi=0$ so that the
constituent quark mass becomes $M_Q^2=g^2(\sigma^2+ {\vec\pi}^2)=g^2\sigma^2=g^2f_{\pi}^2$.
\subsection{Kubo formula, Polyakov Distribution and Transport Coefficients}
In this section, we shall try to write down the expressions for various transport coefficients
using the Green-Kubo formula taking into account the effect of the Polyakov loop.
According to the Green-Kubo relation~\cite{Zubarev,Green,Kubo}, the dissipative
and non-equilibrium quantities like shear viscosity $\eta$, bulk viscosity $\zeta$
and electrical conductivity $\sigma$
can be determined from the thermal fluctuations or thermal correlation
functions - $\langle\pi^{ij}(x)\pi_{ij}(0)\rangle_\beta$,
$\langle{\cal P}(x){\cal P}(0)\rangle_\beta$ and $\langle{J}^i(x){J}_i(0)\rangle_\beta$
respectively, where $\langle ..\rangle_\beta$ stands for thermal average.
The operators for $\eta$ and $\zeta$ can be found from total energy-momentum
tensor $T_{\mn}$: \cite{Nicola}
\bea
\pi^{ij}&\equiv&T^{ij}-g^{ij}T^k_k/3~,
\nn\\
{\cal P}&\equiv&-T^k_k/3 -c_s^2T^{00}~,
\label{Operator}
\eea
where $c_s$ is speed of sound in the medium. 
In general, one can write the transport coefficients ${\cal T}$
in terms of corresponding spectral functions $A_{\cal T}$ as
\be
{\cal T}=I_{\cal T}\lim_{q_0,\vq\rightarrow 0}\frac{A_{\cal T}}{q_0}~,
\label{gen_Transport}
\ee
where
\be
I_{({\cal T}=\eta, \zeta, \sigma)}=\frac{1}{20},~\frac{1}{2},~\frac{1}{6}~,
\ee
and $A_I$ is the Fourier transform of the thermal averaged commutator.
\be
A_{\cal T}=\int d^4xe^{iq\cdot x}\langle [{\cal O}_{\cal T}(x),{\cal O}_{\cal T}(0)]\rangle_\beta,
\ee
with
\be
{\cal O}_{({\cal T}=\eta, \zeta, \sigma)}=\pi^{ij}, {\cal P}, {J}^{i}~.
\ee

\begin{figure}
\begin{center}
\includegraphics[scale=0.5]{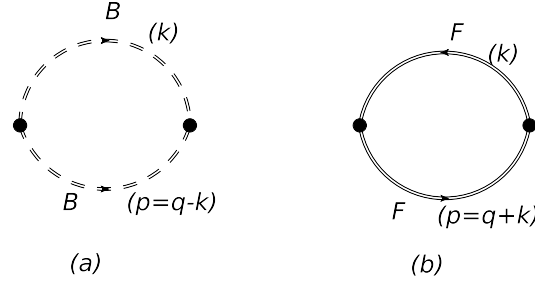}
\caption{One-loop skeleton diagrams for transport coefficients of bosonic (a)
and fermionic (b) medium. Double line of boson-boson (a) and fermion-fermion (b)
internal lines indicate that these propagators are not free propagators,
they contain finite thermal width $\Gamma$.}
\label{one_loop}
\end{center}
\end{figure}

 Using real time thermal field theory the spectral functions can be calculated from the loop diagrams
 as shown in Fig.~\ref{one_loop}(a) and (b) for bosons and fermions respectively.
In real-time thermal field theory, the relation among
the spectral function $A_{\cal T}$, retarded part of self-energy $\Pi^R_{\cal T}$
and $11$-component of self-energy $\Pi^{11}_{\cal T}$ is given by
\be
A_{\cal T}(q)=2{\rm Im}\Pi^R_{\cal T}(q)=2{\rm tanh}(\frac{\beta q_0}{2})
{\rm Im}\Pi^{11}_{\cal T}(q)~.
\label{R_11}
\ee
Evaluating the $11$-component of self-energy and then using Eqs.~(\ref{R_11})
and (\ref{gen_Transport}), the general form of the transport coefficient can
be written as~\cite{G_Kubo,Nicola}
\be
{\cal T}=I_{\cal T}\lim_{q_0,\vq \rightarrow 0}\frac{2}{q_0}
\int\frac{d^3k}{(2\pi)^3}\frac{(-\pi)N_{\cal T}}{4\om_k\om_p}
\{C_2\delta(q_0-\om_k+\om_p)
+C_3\delta(q_0+\om_k-\om_p)\}~,
\label{Tr_G0}
\ee
where $C_2=-f^-_k(\om_k) + f^-_p(-q_0 +\om_k)$,
$C_3=f^+_k(\om_k) - f^+_p(q_0 +\om_k)$ and $f^{\pm}$ are respectively
Fermi-Dirac (FD) distribution functions for particle, anti-particle
of Fermion field.
For bosonic field, $C_2$ and $C_3$ will be interchanged,
although their Bose-Einstein distribution functions for the  particle and the
anti-particle will be same due to vanishing chemical potential.
The energies of intermediate states of the loop diagrams are
$\om_k=\{\vk^2 + m_{\psi,\phi}^2\}^{1/2}$ and
$\om_p=\{|\vq \pm \vk|^2 + m_{\psi,\phi}^2\}^{1/2}$, where $\pm$ stand
for $\psi$ and $\phi$ fields respectively. The quantity $N_{\cal T}$
contains vertex-type factor, which is given by interaction terms in the lagrangian
%

Now the Eq.~(\ref{Tr_G0}) in static limit ($q_0,\vq\rightarrow 0$) gives
a divergent value of ${\cal T}$. This can be easily seen when one uses the identity
\be
\lim_{\Gamma \rightarrow 0}{\rm Im}\left[\frac{-\Gamma}{(q_0 \mp \om_k \pm \om_p)^2 +\Gamma^2} \right]
=(-\pi)\delta(q_0 \mp \om_k \pm \om_p)~,
\ee
so that Eq.~(\ref{Tr_G0}) reduces to
\bea
{\cal T}&=&2I_{\cal T}\lim_{q_0,\vq \rightarrow 0}
\int\frac{d^3k}{(2\pi)^3}\frac{N_{\cal T}}{4\om_k\om_p}
\left\{\frac{(C_2/q_0)(-\Gamma)}{(q_0-\om_k+\om_p)^2+\Gamma^2}
+\frac{(C_3/q_0)(-\Gamma)}{(q_0+\om_k-\om_p)^2+\Gamma^2}\right\}
\nn\\
&=&\frac{I_{\cal T}}{2}\int\frac{d^3k}{(2\pi)^3}
\left(\frac{-N^0_{\cal T}}{\om_k^2\Gamma}\right)
\lim_{q_0,\vq \rightarrow 0}\left(\frac{C_2}{q_0} + \frac{C_3}{q_0} \right )~,
{\rm where}~N^0_{\cal T}=\lim_{q_0,\vq \rightarrow 0}N_{\cal T}
\nn\\
&=&\frac{I_{\cal T}}{2}\int\frac{d^3k}{(2\pi)^3}
\left(\frac{-N^0_{\cal T}}{\om_k^2\Gamma}\right)\beta~F_{\phi,\psi}~,
\label{Tr_G}
\eea
where $\frac{d}{dq_0}(c_2 +c_3)=\beta F$ with
$F=f_k(1 + f_k)$ for $\phi$ and
$f_k(1 - f_k)$ for $\psi$ is
used which clearly diverges in the limit $\Gamma$ $\to$ 0. However for interacting particles
$\Gamma$ will be finite and the delta functions in Eq.~(\ref{Tr_G0}) get replaced by a spectral
function with a finite thermal width.

This adoption of finite thermal width is a very traditional technique
in Kubo framework to get a non-divergent
and finite value of transport coefficients. This $\Gamma$ is inversely related with relaxation
time $\tau=1/\Gamma$.
The relevant vertex like factors are obtained as~\cite{G_Kubo,Nicola,G_CU}
\bea
-N^0_{\eta}&=& \sum_c g\frac{8}{3}\vk^4~,
\nn\\
-N^0_{\zeta}&=& \sum_c 4g\left\{\left(\frac{1}{3}-c_s^2\right)\vk^2
-c_s^2\frac{d}{d\beta^2}(\beta^2m^2_{\psi,\phi})\right\}^2~,
\nn\\
-N^0_{\sigma}&=&4g^e\{\vk^2\}
\eea
and using these in Eq.~(\ref{Tr_G}),
we get expressions of different transport coefficients, as given
in Eqs.~(\ref{shearf}), (\ref{bulkf}) and (\ref{condf}) respectively.
\bea
\eta&=& \sum_c \frac{g \beta}{15}\int\frac{d^3k}{(2\pi)^3}\tau
\left(\frac{\vk^4}{\om_k^2}\right)F~,
\label{shearf}
\\
\zeta&=& \sum_c g \beta \int\frac{d^3k}{(2\pi)^3}\tau
\left(\frac{1}{\om_k^2}\right)
\bigg\{\left(\frac{1}{3}-c_s^2\right)\vk^2
-c_s^2\frac{d}{d\beta^2}(\beta^2m^2)\bigg\}^2 F~,
\label{bulkf}
\eea
\be
\sigma=\sum_c \frac{g^e \beta}{3}\int\frac{d^3k}{(2\pi)^3}\tau
\left(\frac{\vk^2}{\om_k^2}\right)F~.
\label{condf}
\ee
where $g$ is degeneracy factor of medium constituents and
$\tau$ is relaxation time. The degeneracy factor can in general we written as the product of
degeneracies due to spin, color, number of flavors and particle and anti-particle. For example
degeneracy for quark is 2(spin)$\times 2$(flavor)$\times 2$(quark and anti-quark)=8 and sum over the color is still to be taken.
Let us first note that the background gluon field couples to quarks through the covariant derivative as $D_\mu=\partial_\mu-\delta_{\mu 0}A_0$.
In the Polyakov gauge, the Wilson line  $L$ is in the diagonal representation in the color space and therefore, the background gluon field
acts as an imaginary chemical potential for the  colored particles. The corresponding color dependent
equilibrium distribution function for the quarks and the anti-quarks are then given by \cite{pisarskijhep}
\be
f_i(E)=\frac{1}{e^{\beta(E-iQ_i)}+1}; \quad \bar f_i(E)=\frac{1}{e^{\beta(E+iQ_i)}+1}
\label{fqi}
\ee

 where, we have written $A_0^{ij}=\frac{1}{g}\delta^{ij}Q^i$, without any summation over the index $i$. As $A_0$ is traceless, $\sum_iQ^i=0$.
The Polyakov loop $\phi$ is thus related to $ Q_i$ as $\phi=\frac{1}{3}\sum_ie^{i\beta Q_i}$. Further, for vanishing
 baryon density, one can choose $\phi$ to be real and parameterize $Q^i=2\pi T(-q,0,q)$ with $q$ as the dimensionless condensate variable. The Polyakov loop
variable $\Phi$ is therefore given by
\be
\phi=\frac{1}{3}(1+2\cos{2\pi q}).
\label{phiq}
\ee

It is easy to check that the the distribution function of Eq.(\ref{P_FD}) is the color averaged distribution function i.e.
$f_\phi(E)=\frac{1}{3}\sum_i f_i(E)$.

Let us note that in Eq.(\ref{Tr_G}) the summation over all the colors for the fermion loops needs to be done. Thus while summing over colors one has
\bea
F &\equiv&
\sum_c f_c(\om_k)\{1-f_c(\om_k)\}
\nn\\
&=&3 f_\Phi -\frac{3}{D^2}\bigg[e^{-6\beta\om_k}+
\Phi(3\Phi-2)e^{-2\beta\om_k}
+4\Phi e^{-4\beta\om_k}
+2\Phi (3\Phi-1)e^{-3\beta\om_k}\bigg],
\nn\\
\label{f2}
\eea
where, $D$ is the denominator of the Polyakov loop distribution function (\ref{P_FD}),
$D=1+3 \Phi e^{-\beta \om_k}+3 \Phi e^{-2\beta \om_k}+3 e^{-3\beta \om_k}$.
%
Let us note that one might further approximate the color dependent distribution functions by their color averaged distribution function of Eq.(\ref{P_FD}) in which case
\begin{equation}
 F \simeq 3f_\phi (1-f_\phi)
\end{equation}
In such cases e.g., the expression for $\eta$ becomes the more familiar expression as
\bea
\eta&=&  \frac{g N_c \beta}{15}\int\frac{d^3k}{(2\pi)^3}\tau
\left(\frac{\vk^4}{\om_k^2}\right) f_\phi(1-f_\phi)~,
\label{shearf2}
\eea
One may note that such a replacement of color averaged distribution function is exact in the Boltzmann limit.
Further difference between replacing the color distribution functions and their color averaged one is
proportional to $\phi (\phi -1) e^{-2\beta E}$. This difference is small both below and above the critical
temperature while it can be relevant around the critical temperature.
%

So far we have been concerned with the transport coefficients of quark
components, where we have to take care of FD distribution with
color degrees of freedom, which ultimately gives the Polyakov
loop distribution function. However, for meson components, we
have to use just BE distribution function in the $F$ of
Eqs.~(\ref{shearf}), (\ref{bulkf}) and (\ref{condf}).

\begin{figure}
\centering
\includegraphics[scale=0.82]{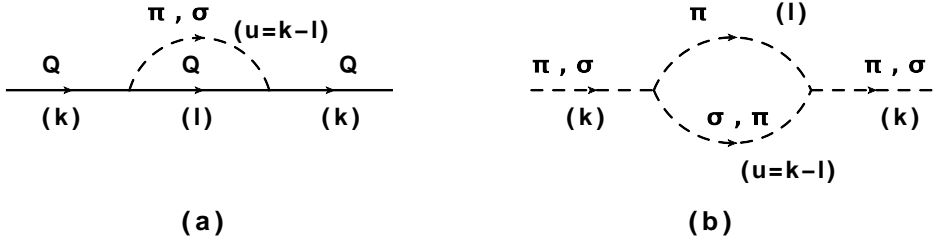}
\caption{Diagram {\bf (a)} represents quark self-energy diagrams with quark-meson
loop, where $M=\pi ~,~ \sigma$. Diagram {\bf (b)} denotes $\pi$ and $\sigma$ meson
self-energy diagram for $\pi$-$\sigma$ and $\pi$-$\pi$ loop respectively. }
\label{PQM_loop}
\end{figure}
%
%
%

\subsection{Thermal widths of quarks and mesons}
\label{widthsub}
We have calculated the thermal widths of different components from the imaginary part of their self energies.
Unlike the vacuum case, for finite temperature, these thermal widths have contributions from both the decay processes and scattering
processes involving particles present in the medium. Depending on the mass of the component we are interested in, either of
the processes can become dominant. In our calculations, we found that for sigma meson, the mass is such that decay process dominates whereas,
for pion and quark cases, the scattering is the dominant process. Thus
to calculate the thermal width of quarks and mesons, we have to use relevant
interaction part from the total Lagrangian density, addressed in Eq.~(\ref{lagpqm}).
From Eq.~(\ref{uchi}), we can identify $\sigma\pi\pi$ interaction Lagrangian
density:
\be
{\cal L}_{\sigma\pi\pi}= \lambda f_\pi\sigma \vec\pi^2~,
\label{LSpipi}
\ee
which will help
us to calculate mesonic thermal widths $\Gamma_M$. Similarly to calculate
quark thermal widths $\Gamma_Q$, we need to identify
the $QQM$ interaction part from Eq.~(\ref{lagpqm}). Expanding this $QQM$ interaction
Lagrangian density and dividing it into $QQ\pi$ and $QQ\sigma$ components, we get
\be
{\cal L}_{QQ\pi}=ig\left[\sum_{Q=u,d}\bar{\psi_Q}\gamma^5\pi^0\psi_Q
+\sqrt{2}\{\bar{\psi_u}\gamma^5\pi^+\psi_d+~{\rm h.c.}\}\right]
\label{LQQpi}
\ee
and
\be
{\cal L}_{QQ\sigma}=g\sum_{Q=u,d}\bar{\psi_Q}\sigma\psi_Q
\label{LQQsig}
\ee
respectively.

The quark thermal width, $\Gamma_Q$, can be estimated from the retarded
part of the quark self-energy $\Sigma^R_{Q(QM)}$ at finite temperature for
quark-meson $(QM)$ loops, where $M=\pi,~\sigma$
as shown in Fig~\ref{PQM_loop}(a). With the help of interaction
Lagrangian densities (\ref{LQQpi}) and (\ref{LQQsig}), we  obtain
\bea
\Gamma_Q (\vk)&=&-\sum_{M=\pi,\sigma}
\left[{\rm Tr}\left\{\frac{(\ks +M_Q)}{2M_Q}{\rm Im}{\Sigma}^R_{Q(QM)}(k)
\right\}\right]_{k_0=\om^k_Q}~
\nn\\
&=&\left[\int\frac{d^3{\vec l}}{(2\pi)^3}
\{n_Q(\om^l_Q)+n_\pi(\om^u_\pi)\}\delta(k_0+\om^l_Q-\om^u_\pi)
\right.\nn\\
&&\left.\frac{3g^2{\rm Tr}\left[(\ks +M_Q)
\gamma^5(\ls +M_Q)\gamma^5\right]}{2M_Q(4\om^l_Q\om^u_\pi)}\right]_{l_0=-\om^l_Q,k_0=\om^k_Q}
\nn\\
&&+\left[\int\frac{d^3{\vec l}}{(2\pi)^3}
\{n_Q(\om^l_Q)+n_\sigma(\om^u_\sigma)\}\delta(k_0+\om^l_Q-\om^u_\sigma)
\right.\nn\\
&&\left.\frac{g^2{\rm Tr}\left[(\ks +M_Q)
(\ls +M_Q)\right]}{2M_Q(4\om^l_Q\om^u_\sigma)}\right]_{l_0=-\om^l_Q,k_0=\om^k_Q}~,
\label{g_Q}
\eea
%
where $n_Q(\om^l_Q)$
and $n_{\pi,\sigma}(\om^u_{\pi,\sigma})$
are the FD and BE distribution functions for
intermediate $Q$ and $M=\pi,\sigma$ states
respectively. Here, we can again replace $n_Q(\om^l_Q)$ by color average
thermal distribution or Polyakov loop distribution $f_\Phi$.
Now as mentioned earlier the relevant process for quark is the QQ$\pi$ or QQ$\sigma$ scattering processes,
analyzing the detailed branch cuts of this quark self-energy
at finite temperature, one should notice that the quark
pole ($k_0=\om^k_Q,\vk$) remains within the Landau-cut (cut corresponding to the scattering process)
region( ($\vk <k_0<\{\vk^2+(M_Q-M_\pi)^2\}^{1/2}$) of $Q\pi$ loop, when
$M_\pi>2M_Q$. Therefore, we will get non-zero values of $\Gamma_Q$
only in the temperature region, where $M_\pi>2M_Q$.
This Landau cut contribution of quark self-energy basically interprets forward and backward
quark-meson scattering, by which mesons are absorbed and emitted respectively~\cite{G_IFT}.

Similarly, thermal widths of pion and sigma mesons
$\Gamma_\pi$ and $\Gamma_\sigma$ can be estimated from the retarded
part of the meson self-energies $\Pi^R_{\pi(\pi\sigma)}$ and
$\Pi^R_{\sigma(\pi\pi)}$ respectively, which are
represented by Fig~\ref{PQM_loop}(b) in a general form. Using the interaction
Lagrangian density (\ref{LSpipi}), they are respectively derived as
\bea
\Gamma_\pi (\vk)&=&-\frac{1}{M_\pi}[{\rm Im}{\Pi}^R_{\pi(\pi\sigma)}(k)]_{k_0=\om^k_\pi}
\nn\\
&=&\left[\int\frac{d^3{\vec l}}{(2\pi)^3}
\{n_\pi(\om^l_\pi)-n_\sigma(\om^u_\sigma)\}
\delta(k_0+\om^l_\pi-\om^u_\sigma)
\left(\frac{\lambda^2 f_\pi^2}{M_\pi}\right)
\frac{1}{4\om^l_\pi\om^u_\sigma}
\right]_{k_0=\om^k_\pi}~
\nn\\
\label{g_p}
\eea
and
\bea
\Gamma_\sigma (\vk)&=&-\frac{1}{M_\sigma}[{\rm Im}{\Pi}^R_{\sigma(\pi\pi)}(k)]_{k_0=\om^k_\sigma}
\nn\\
&=&\left[\int\frac{d^3{\vec l}}{(2\pi)^3}
\{1 + n_\pi(\om^l_\pi)+ n_\pi(\om^u_\pi)\}\delta(k_0-\om^l_\pi-\om^u_\pi)
\left(\frac{\lambda^2 f_\pi^2}{M_\sigma}\right)
\frac{1}{4\om^l_\pi\om^u_\pi}\right]_{k_0=\om^k_\sigma}~,
\nn\\
\label{g_s}
\eea
where $n_\pi$ and $n_\sigma$ are BE
distribution functions for intermediate $\pi$ and $\sigma$ states respectively.

Analyzing the detailed branch cuts of  pion and sigma meson self-energies
at finite temperature, one can find that the pion
pole ($k_0=\om^k_\pi,\vk$) and sigma pole ($k_0=\om^k_\sigma,\vk$)
are respectively situated in the Landau-cut ($\vk <k_0<\{\vk^2+(M_\sigma-M_\pi)^2\}^{1/2}$)
and unitary-cut ($\{\vk^2+4M_\pi^2\}^{1/2} <k_0< \infty$) regions
for certain temperature range, where $M_\sigma > 2M_\pi$.
Here, Landau cut contribution of pion self-energy measures the probabilities of
forward and backward $\pi$-$\sigma$ scattering, where a $\sigma$ is absorbed by
former process and emitted by latter one. Next, the unitary cut contribution of
$\sigma$ meson self-energy signifies forward and backward decay processes -
$\sigma\rightarrow \pi\pi$ and $\pi\pi \rightarrow \sigma$ respectively.
Unlike the earlier Landau cuts for quark and pion self-energies, the unitary
cuts of $\sigma$ meson self-energy remains non-vanishing at $T=0$ as it is associated
with forward decay process. At finite temperature, this unitary cut contribution gives
a Bose-enhanced probability of this forward decay process and also an in-medium probability
of backward decay process, which is absent in vacuum. 
\section{Numerical results and discussion}
\begin{figure}
\centering
\includegraphics[scale=0.35]{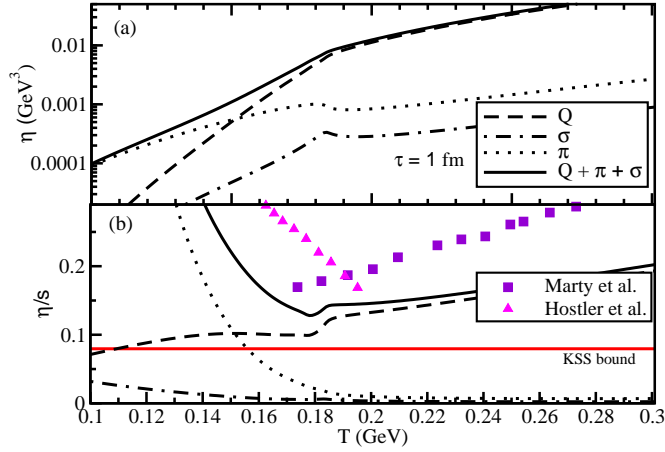}
\caption{(Color online) Temperature dependence of $\eta$ (a) and $\eta/s$ (b) for quark (dashed line),
pion (dotted line) and sigma (dashed-dotted line) components
for constant interaction strength.
The results of Marty et al.~\cite{Marty} (brown circles),
Fraile et al.~\cite{Nicola} (magenta stars), Lang et al.~\cite{Weise1} (green squares)
are included for comparison.
The straight horizontal red line indicates the KSS bound.}
\label{eta_T}
\end{figure}

Let us start our numerical investigations by using the expressions
of $\eta$, $\zeta$ and $\sigma$, given in Eqs.~(\ref{shearf}$-$\ref{condf}).
At first, instead of using explicit temperature ($T$) and momentum ($\vk$)
dependent thermal width of quarks and mesons, constant values will be
considered to highlight the thermodynamical phase space structure of
different transport coefficients. This study is important
as strong $T$ dependence of thermal width sometimes dominates over the $T$
dependence of phase-space part. For example, we have conformal symmetry breaking
term in phase-space part of bulk viscosity expression (\ref{bulkf}), which
generally gives a peak structure near the transition temperature. Hence,
bulk viscosity able to map that peak structure only
for a mild $T$ dependent or constant values of thermal width of medium constituents
as we notice in Refs.~\cite{Kinkar_bulk,De-Fu,Dobado_zeta1,Dobado_zeta2,G_IFT}.
However, a strong $T$ dependent thermal width or relaxation time can suppress
the foot print of peak structure as observed in Refs.~\cite{Deb,Marty,Redlich_NPA,G_IFT}.

The results of $\eta(T)$ for
constant $\Gamma_Q$, $\Gamma_\pi$ and $\Gamma_\sigma$ are plotted
by dashed, dotted and dash-dotted lines in Fig.~\ref{eta_T}(a).
One can define their respective relaxation times
$\tau_Q$, $\tau_\pi$ and $\tau_\sigma$ using the relation
$\tau_{(Q,\pi,\sigma)}=1/\Gamma_{(Q,\pi,\sigma)}$ and
we have fixed their values to 1 fm as a typical value. The total
shear viscosity $\eta_t=\eta_Q +\eta_\pi +\eta_\sigma$ is denoted by
solid line in Fig.~\ref{eta_T}(a), which shows the dominant contribution
of the quark component, compared to the contributions of the components $\pi$ and $\sigma$
mesons. $\eta(T)$ of all components and their total appear as increasing
functions of $T$ but they exhibit some changes in their rate of increase near the
transition temperature $T_c$. Here, we assume that the interactions are well localized
in space and time, so that total energy-momentum tensor, electromagnetic current
for the multi-component system can be
approximated as a sum of independent contributions~\cite{Purnendu}. Therefore,
total transport coefficient is just considered as summation of individual
components.

The Fig.~\ref{eta_T}(b) shows the variation for $\eta/s$ with temperature
for respective components and their total,
where we notice that mesonic components decrease and quark component
increases with temperature.
%
%
%
The increasing trend of total $\eta/s$ in the quark temperature domain
is supported by the results of Ghosh et al.~\cite{G_IFT} (cyan triangles)
and Marty et al.~\cite{Marty} (violet squares).
The decreasing trend of the ratio below the transition temperature is also
in agreement with the standard hadronic model
calculations~\cite{Itakura,Dobado,Nicola,Weise1,SSS,Ghosh_piN,Gorenstein,HM,HM2,Hostler}.
Total $\eta/s$ is above the so-called KSS bound~\cite{KSS}.

\begin{figure}
\centering
\includegraphics[scale=0.35]{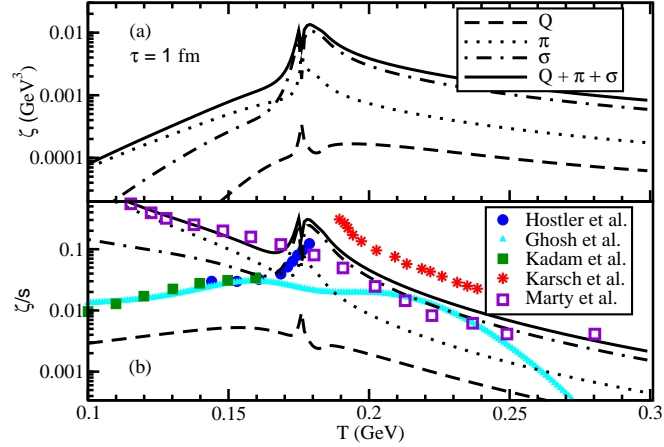}
\caption{(Color online) Temperature dependence of $\zeta$ (a) and
$\zeta/s$ (b) of quark (dashed line),
pion (dotted line), $\sigma$ (dash-dotted line) and their total (solid line)
for constant interaction strength.
The results are compared with Karsch et al.~\cite{Tuchin2} (red stars),
Hostler et al.~\cite{Hostler} (blue circles) and Marty et al.~\cite{Marty} (open squares)}. 
\label{zeta_T}
\end{figure}

Following same notations of curves in Fig.~(\ref{eta_T}),
the temperature dependence of $\zeta$ and $\zeta/s$
are shown in Fig.~\ref{zeta_T}(a) and (b) respectively.
Here, we are observing a sharp peak structure in
$\zeta$ and $\zeta/s$ near $T_c$. The peak structures of
pion and sigma components dominate compared to that
of quark component.
To understand this, we have to
focus on the conformal symmetry breaking terms~\cite{Purnendu,G_IFT}
$\left(\frac{1}{3} - c_s^2\right)$ and
$\frac{d}{d\beta^2} \left( \beta^2M^2 \right)$ in the
integrand of Eq.~(\ref{bulkf}), where $M$ is the temperature dependent mass
of medium constituent.
Near $T_c$, the contribution
of these quantities become maximum, which is at the root of
this peak structure in $\zeta$. Hence, one can associate this
peak structure of $\zeta(T)$ with the maximum violation of conformal
symmetry breaking near $T_c$. The first term of conformal symmetry breaking
is same for all components, while second one is different for three different
components. Rate of change of their masses with temperature reach their extrema
near $T_c$, which is responsible for making peaks in their $\zeta$'s. Interestingly,
sigma meson component shows two peak structure because of the non-monotonic changes
of its mass with temperature. Such a peak structure has also been observed in the earlier calculations,
based on Linear Sigma Model~\cite{Purnendu,Dobado_zeta2}
and Nambu-Jona-Lasinio model~\cite{G_IFT}, where similar kind of
peak structure in $\zeta$ near $T_c$ were observed.
We may assume indication of similar kind
of peak structure from the increasing nature
of $\zeta/s(T<T_c)$~\cite{HM,Hostler} in the hadronic
temperature range and the decreasing nature of
$\zeta/s(T>T_c)$~\cite{Meyer_zeta,Tuchin,Tuchin2} in the temperature region
of quark phase. Fig.~\ref{zeta_T}(b) has included some earlier
results of $\zeta/s$ by
Ghosh et al.~\cite{G_IFT} (triangles), Kadam et al.~\cite{HM} (solid squares),
Karsch et al.~\cite{Tuchin2} (stars), Hostler et al.~\cite{Hostler} (blue circles),
Marty et al.~\cite{Marty} (open squares),
whose order of magnitude is roughly close to the present estimations.

\begin{figure}
\centering
\includegraphics[scale=0.35]{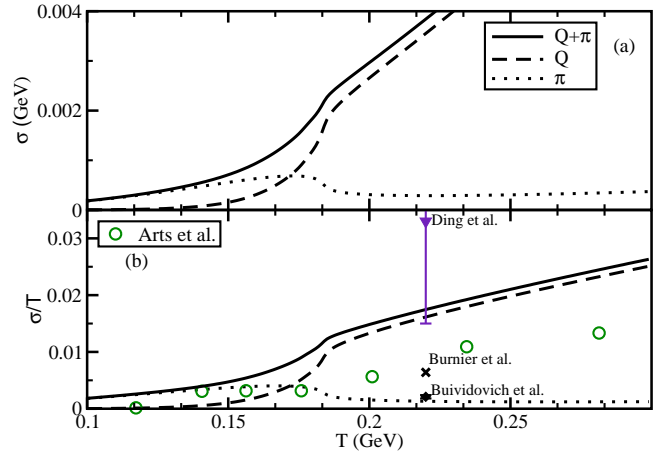}
\caption{(Color online) Temperature dependence of $\sigma$ (a) and $\sigma/T$ (b)
for quark (dash line) and pion (dotted line) components for constant interaction strength
and they are compared with earlier results by Fraile et al.~\cite{Nicola} (magenta stars), 
Ding et al.~\cite{LQCD_Ding} (triangle down) Aarts et al.~\cite{LQCD_Arts_2015} (open circle),
Burnier et al.~\cite{LQCD_Burnier} (cross) and Buividovich et al.~\cite{LQCD_Buividovich} (diamond).}
\label{el_T}
\end{figure}

Next, Fig.~\ref{el_T}(a) and (b)  respectively show the
electrical conductivity $\sigma$ and the dimensionless quantity $\sigma/T$
for different components of the medium. Using Eqs.~(\ref{condf}) for different
constituents, the results for quark (dashed line), pion (dotted line) components
and their total (solid line) are calculated, where charge neutral
constituents like $\sigma$ and $\pi^0$ do not come into the picture.
Similar to shear viscosity, $\sigma(T)$ of different components and their total are
increasing functions of $T$ with some changes in
their rate of increments near the transition temperature $T_c$.
The $\sigma(T)$ of quark component is much larger than that of
pion component at high $T$ domain but in low temperature domain,
pion component is dominant over quark component.
In the hadronic temperature
domain, $\sigma/T$ of Refs.~\cite{Lee,Nicola_PRD,Finazzo} are more or less
in the same order of magnitude as our result. In this context, there are large numbers of
works in LQCD approach with different numerical strengths of $\sigma/T$.
Some of them~\cite{LQCD_Buividovich,LQCD_Burnier,LQCD_Ding} are
displayed in Fig.~\ref{el_T}(b) at certain $T$ ($>T_c$). A temperature dependent
data (open circles) for $\sigma/T$ is also added from the latest version of
Ref.~\cite{LQCD_Arts_2007}.
\begin{figure}
\centering
\includegraphics[scale=0.35]{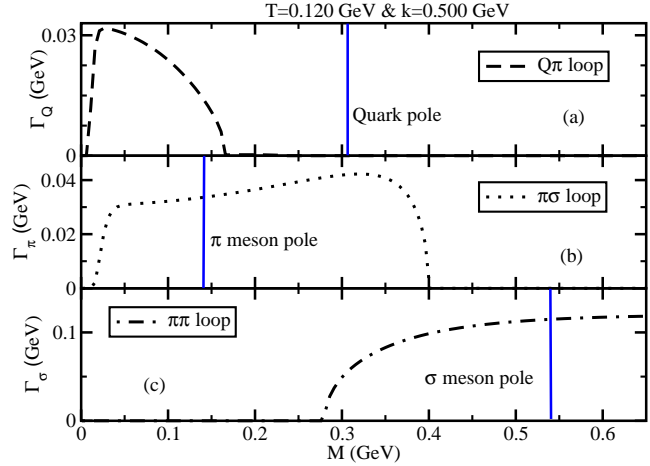}
\caption{(Color online) Invariant mass distribution of thermal
width of quark (a), pion (b) and sigma meson (c)
at $T=0.120$ GeV and $|\vk| =0.500$ GeV.
The vertical blue lines in each panel correspond
to the pole mass of the respective particles.}
\label{G_M}
\end{figure}
\begin{figure}
\centering
\includegraphics[scale=0.35]{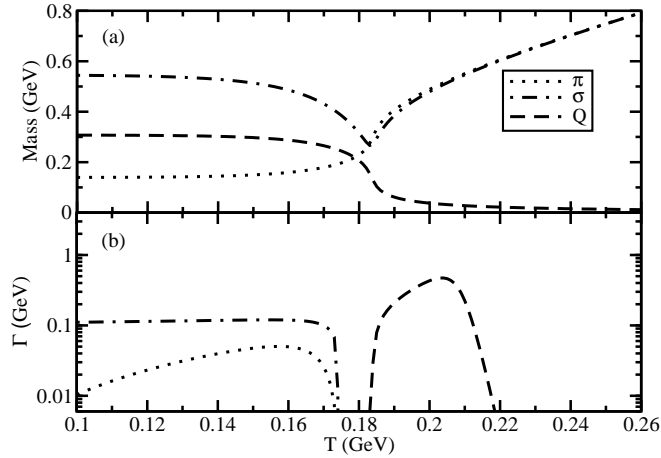}
\caption{Temperature dependence of mass (a) and thermal width (b)
for quark (dash line), pion (dotted line) and sigma meson (dash-dotted line).
Momentum is fixed at $|\vk|=0.500$ GeV for panel (b).}
\label{MG_T}
\end{figure}

So far, the results of transport coefficients $\eta$, $\zeta$ and $\sigma$
are obtained for constant thermal widths of medium constituents to highlight the phase-space structure of these coefficients.
Now we proceed to estimate these transport coefficients for explicit
$T$ and $\vk$ dependent thermal width.
Before that, let us discuss in detail the structure of thermal widths for different
medium constituents and then the results of different transport coefficients
using those temperature and momentum dependent $\Gamma$'s.
For quark thermal width $\Gamma_Q$, let us first concentrate on Eq.~(\ref{g_Q}),
which gives the on-shell value ($k_0=\om_Q$) of the imaginary part of quark
self-energy for quark-meson loops. We can see the invariant mass distribution
of $\Gamma_Q$ by transforming the on-shell relation $k_0=\om^k_Q=\{\vk^2 + M_Q^2\}^{1/2}$
to the off-shell one $k_0=\{\vk^2 + M^2\}^{1/2}$
in Eq.~(\ref{g_Q}), where $M$ is invariant mass of quark.
Now we can expect to observe a scattering (pion, quark) or decay (sigma) interaction at a particular temperature
if the pole mass of the constituent lies within the range, where the invariant mass distribution of the thermal width is nonzero.
For temperature $T=0.120$ GeV and quark
momentum $\vk=0.500$ GeV, Fig.~\ref{G_M}(a) shows 
the structure of $\Gamma_Q(M)$ (dashed line). Here the straight vertical blue line
denotes the on-shell (constituent) quark mass $M_Q$ at $T=0.120$ GeV, obtained
within PQM model. We can see that the $M_Q$ is away from the Landau
cut region ($0<M<|M_Q-M_\pi|$) of quark self-energy, where $\Gamma_Q(M)$ is non-zero.
Therefore, the on-shell value of $\Gamma_Q$ at $T=0.120$ GeV is zero, which we can see
from the dashed line of Fig.~\ref{MG_T}(b). Remembering the discussion related to this
issue after Eq.~(\ref{g_Q}), we can get a non-zero on-shell value of $\Gamma_Q$
beyond the Mott temperature $T_M$, from where the threshold condition
$(M_\pi>2M_Q)$ of $\pi QQ$ interaction will be valid.
In other words, we can say that $M_Q$ will
be within the Landau cut region of quark self-energy for $T>T_M$.

Imposing similar kind of off-shell condition in Eqs.~(\ref{g_p}) and
(\ref{g_s}) for $\pi$ and $\sigma$ mesons, one obtains
$\Gamma_\pi(M)$ and $\Gamma_\sigma(M)$ which are shown in Fig.~\ref{G_M}(b) and
(c) respectively. We see that the on-shell masses of $\pi$ and $\sigma$ mesons at
$T=0.120$ GeV, indicated by blue vertical lines in Fig.~\ref{G_M}(b) and
(c), are located within the respective branch cuts of their self-energies.
Therefore, we will get non-zero on-shell values of $\Gamma_\pi$ and $\Gamma_\sigma$
at $T=0.120$ GeV. After a certain temperature, from where the threshold condition
$(M_\sigma > 2M_\pi)$ of $\sigma\pi\pi$ interaction is not valid,
the on-shell values of $\Gamma_\pi$ and
$\Gamma_\sigma$ will vanish. This fact becomes more clear in Fig.~\ref{MG_T}(b),
where on-shell values of $\Gamma_\pi$ (dotted line), $\Gamma_\sigma$ (dash-dotted line)
and $\Gamma_Q$ (dashed line) are plotted against $T$ axis.
Now following our
previous discussion in section \ref{widthsub}, we see that $M_Q$ is only smaller
than $M_{\sigma} - M_Q$ or $M_{\pi}-M_Q$ for $T>T_M$ (Mott  temperature)
and it is never greater than the sum of the masses of any two components. So as we mentioned
earlier it will only go through scattering processes and
that is also only for a limited temperature region which we can clearly see in the lower panel,
where $\Gamma_Q$ remains zero in hadronic temperature
region. Beyond the Mott temperature, it gets the non-zero value. Similarly we notice that $\Gamma_\pi$
and $\Gamma_\sigma$ are non-zero in hadronic temperature range and beyond the Mott
temperature, they vanish. These $T$ dependence of thermal widths are mainly controlled by the $T$
dependence of $M_Q$, $M_\pi$ and $M_\sigma$, which
are shown by dashed, dotted and dash-dotted lines respectively in Fig.~\ref{MG_T}(a).
The masses of $\sigma$ and $\pi$ mesons are basically obtained from the
Eqs.~(\ref{sigmass}) and (\ref{pimass}).
In the chirally broken phase, the pion mass, being the mass
of an approximate Goldstone mode, is protected and varies weakly with temperature. On the other hand,
the mass of $\sigma$, which is approximately twice of the constituent quark mass, drops significantly near the
transition temperature. At high temperature, being chiral partners, the masses of $\sigma$ and $\pi$
mesons become degenerate and increase linearly with temperature.
The temperature dependence of quark mass $M_Q=g~\sigma$ is mainly determined by the temperature
dependence of chiral order parameter $\sigma$, which decreases
with temperature to small values but never vanishes.
On the other hand, the Polyakov loop parameter 
grows from $\Phi(T=0)=0$ to $\Phi=1$ at high temperatures. This decreasing
and increasing nature of ${\sigma}(T)$ and $\Phi(T)$ respectively signify the chiral and confinement properties of quark-hadron phase transition and are reflected in the temperature dependence of quark and meson masses. The transport coefficients
thus depend both on the phase space factors and the momentum dependent widths.
\begin{figure}
\centering
\includegraphics[scale=0.35]{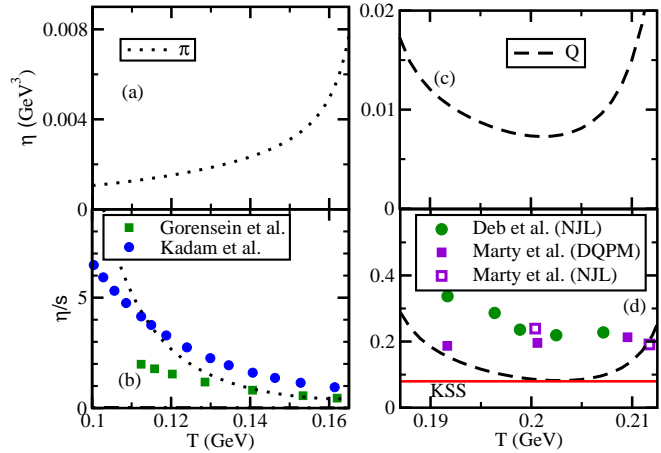}
\caption{(Color online) Temperature dependence of
$\eta$ (a) and $\eta/s$ (b)
for pion (dotted line) and quark (dashed line)
using their temperature and momentum dependent
thermal widths and comparison with the results of
Kadam et al.~\cite{HM2} (blue circles), Gorenstein et al.~\cite{Gorenstein}
(green squares),
Deb et al.~\cite{Deb} (green circles), Marty et al.~\cite{Marty} (solid and open squares
for DQPM and NJL model).}
\label{eta_T2}
\end{figure}
\begin{figure}
\centering
\includegraphics[scale=0.35]{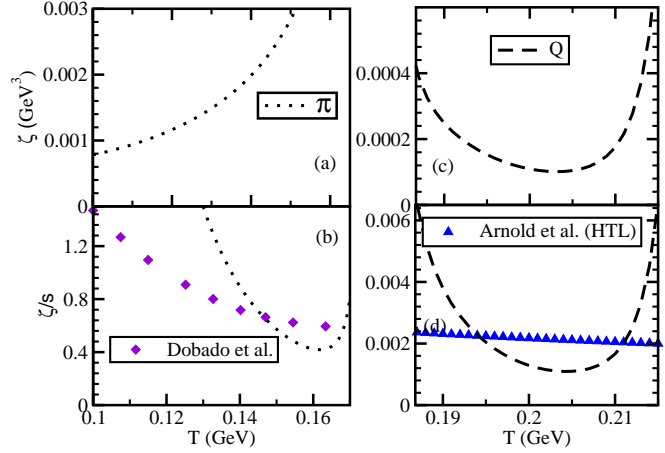}
\caption{(Color online) Temperature dependence of $\zeta$ (a) and
$\zeta/s$ (b) of quark (dashed line)
pion (dotted line) by using their temperature and momentum dependent
thermal widths and compared with the results of
Dobado et al.~\cite{Dobado_zeta2} (violet diamonds),
Arnold et al.~\cite{Arnold_bulk} (blue triangles).}
\label{bulk}
\end{figure}
\begin{figure}
\centering
\includegraphics[scale=0.35]{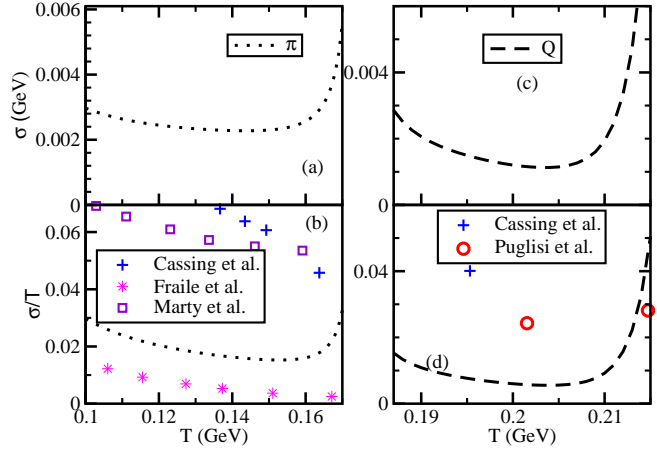}
\caption{(Color online) Temperature dependence of
$\sigma$ (a) and $\sigma/T$ (b)
for pion (dotted line) and quark (dashed line)
using their temperature and momentum dependent
thermal widths and comparison with the results of
Fraile et al.~\cite{Nicola} (magenta stars), Marty et al.~\cite{Marty}
(open squares),
Cassing et al.~\cite{Cassing} (blue pluses),
Puglisi et al.~\cite{Puglisi} (open circles).}
\label{el_T3}
\end{figure}

Now, when we use the explicit
structure of $\Gamma_{(Q,\pi,\sigma)}$ in Eqs.~(\ref{g_Q}), (\ref{g_p})
and (\ref{g_s}), the Figs.~(\ref{eta_T}), (\ref{zeta_T}) and (\ref{el_T})
are respectively transformed to Figs.~(\ref{eta_T2}),
(\ref{bulk}) and (\ref{el_T3}), which have now additional influence of
temperature dependent thermal widths along with the phase space part.
We have split the temperature region into
two parts because of the kinematic thresholds for mesons and quark self-energies.
The first part, covering the hadronic temperature range, is for the results of
pion component, while the results of quark component are plotted in the second
part, which can be considered as quark temperature region. Since the results of
$\sigma$ meson component is negligible, because of its high values of
$\Gamma_\sigma$ as well as $M_\sigma$ in the hadronic temperature range, the
results of pion component only are shown in the panel of hadronic
temperature region. Let us first come to the Fig.~\ref{eta_T2}, where
$\eta$ and $\eta/s$ for pion and quark components are described
in panels (a), (b), (c) and (d) respectively. The blowing up tendency
of $\eta$ for pion component in high $T$ domain is because of the suppressing
tendency of $\Gamma_\pi$, as shown previously by the dotted line in Fig.~\ref{MG_T}(b).
Similarly, the result of $\eta$ for quark component is also limited within a particular
(high) temperature domain because the non-zero structure of $\Gamma_Q(T)$ is limited
within that $T$-zone, as shown previously by the dashed line in Fig.~\ref{MG_T}(b).
That is why we should focus on the order of magnitude for the transport coefficients
instead of those blowing up regions. By using this kind of $\Gamma(\vk,T)$, not only
$\eta$ but also other transport coefficients follow this pattern, which also observed
in the earlier Ref.~\cite{G_IFT}.
Fig.~\ref{eta_T2}(b) shows that
 the $\eta/s$ for pion component from our calculation has, approximately, same order of magnitude
as the results of Kadam et al.~\cite{HM2} (blue circles), Gorenstein et al.~\cite{Gorenstein}
(green squares).
The corresponding results for quark component in its $T$ range agree with the earlier
results obtained by Deb et al.~\cite{Deb} (green circles) and
Marty et al.~\cite{Marty} (solid and open squares for two different models).
%

Following similar pattern of Fig.~(\ref{eta_T2}), Fig.\ref{bulk}(a) to (d)
show the $\zeta$ and $\zeta/s$ for pion and quark components,
whose order of magnitude are comparable with results of
Dobado et al.~\cite{Dobado_zeta2} (violet diamonds)
in the hadronic temperature range and
Arnold et al.~\cite{Arnold_bulk,Kapusta:2008vb} (blue triangles) in
the quark temperature domain.
Next, using the same $T$ and $\vk$ dependent thermal widths, the electrical
conductivity $\sigma$ and $\sigma/T$ for pion and quark components are plotted
in Fig.~(\ref{el_T3}). Numerical values of $\sigma/T$, obtained by us, are compared
with the results of  Marty et al.~\cite{Marty}
(open squares), Cassing et al.~\cite{Cassing} (blue pluses),
Fraile et al.~\cite{Nicola_PRD,Nicola} (magenta stars) in the hadronic temperature domain,
whereas in the quark temperature domain, the values of our $\sigma/T$ are compared
with the results of
Cassing et al.~\cite{Cassing} (blue pluses) and Puglisi et al.~\cite{Puglisi} (open circles).


One may note that the present estimation of the transport coefficients in divided into two narrow temperature intervals about the Mott temperature.
Below the Mott temperature, one has contributions from hadrons, namely pions and sigma while above the Mott temperature one has contributions from quarks.
Below the Mott temperature the dominant contribution to transport coefficients comes from pions. The kinematic constriant of m$_\pi < |m_\pi-m_\sigma |$
translates into an upper cutoff on temperature for this process. Similary above the Mott temperature the constraint m$_\pi>2 m_q$ imposes a lower cut off
on temperature. As shown in Fig. 7.b., $\Gamma_q$ vanishes for large temperatures. Thus the 1$\rightarrow$2 processes considered in this work results in
the two narrow intervals near Mott temperature. However the temperature ranges of these intervals can be extended by including off shell contribution as
has been considered in Ref~\cite{Weise3}. Besides the calculation can be improved by including 2$\rightarrow$2 scattering process which will not be
restricted by such kinematic constriants \cite{HM_PQM2}.

One may observe from this investigation that near (but above)
the Mott temperature $Q\leftrightarrow QM$ scattering plays an important role
as including these interactions result in a ratio of $\eta/s$ close to the conjectured KSS lower bound of $1/(4\pi)$. In the temperature interval above Mott temperature, the divergent nature of the transport coefficients at high temperatures reflect the fact that $Q\leftrightarrow QM$
scattering contribution to $\Gamma_q$ vanishes. Although, one can expect an approximately
constant values of $\eta/s$ ($\sim 0.3)$ if one considers the $2\rightarrow 2$ elastic
scattering of quark-quark and quark-meson interaction as discussed in Ref.~\cite{HM_PQM2}.

We might realize here that a small value of $\eta /s$ near the transition temperature is a reflection of an inherent non-perturbative feature
captured by effective models like NJL \cite{Redlich_NPA,Marty,Deb} and PQM \cite{HM_PQM2} models unlike pQCD calculations at high temperature \cite{Arnold}.
Such models consider $2\rightarrow 2$ elastic scatterings. The inclusion of in-elastic scattering component with those earlier estimations will lead a
further reduction of $\eta/s$ near the Mott temperature.
Similar effects are also expected in other transport coefficients like bulk viscosity and electrical conductivity.

The present approximation of estimating the transport coefficients is similar to quasi-particle  and relaxation
time approximation of the boltzmann equation as done in Ref~\cite{Purnendu}. Within this approximation the contritbution of
different species for transport coefficients are added. Whereas to calculate the relaxation time for given species the relaxation
times of scattering and decay involving other particles are added inversely. Therefore the effect of interactions with other species
is also included in the evaluation of relaxation time and hence on the transport coefficient.

Besides such RTA mixing picture, there are also some alternative prescriptions like Ref~\cite{denicol2} which is similar to Chapman-Enskog method,
approximate method of Wilke's expression~\cite{Wilke} and many empirical based methods~\cite{Saxena}.

%
\section{Summary and Perspectives}
In this article, we have investigated the role of PQM dynamics in calculations
of different transport coefficients like shear viscosity $\eta$, bulk viscosity $\zeta$
and electrical conductivity $\sigma$ of quark and hadronic medium with zero
quark chemical potential.
We have first briefly sketched the background framework of Kubo method to
obtain the standard expressions of $\eta$, $\zeta$ and $\sigma$.
We have also shown explicitly how the polyakov loop affects the thermal distribution functions of quarks which appear in calculation of transport coefficients.

The temperature dependence of transport
coefficients depends on two parts - one is thermodynamical phase-space structure
and other is thermal width of respective transport correlators. To distinguish
between the two effects, we have taken constant thermal width for different transport correlators,
which essentially reveals the effect of the phase-space structure on different transport coefficients.
The phase-space structure of transport coefficients is mostly governed by the temperature
dependent masses of the quark and mesons in PQM model. The temperature dependence of the masses is
governed by the chiral symmetry restoration in the model at high temperature.
$\eta$ and $\sigma$ show an increasing behavior with temperature without much sigmnificant changes at the quark-hadron transition temperature. 
 
Interestingly, even with constant decay width $\eta/s$ shows a minimum at the tranisition temperature. 

On the other hand $\zeta$ exhibits a peak structure near the transition temperature. Unlike $\eta$
and $\sigma$, the expression of $\zeta$ additionally contains
conformal breaking terms of QCD medium apart from phase space structure, whose peak pattern is reflected in temperature dependence of $\zeta$.
The qualitative profiles of $\eta/s$, $\zeta/s$ and $\sigma/T$ vs $T$ are in agreement
with the earlier studies, based on different dynamical models.

Next, we have considered the explicit temperature and momentum dependence of $\Gamma$ for quark
and mesons to estimate the values of those transport coefficients.
In our earlier work~\cite{HM_PQM2}, the quark and pion thermal widths or their inverse (relaxation times)
have been calculated using $2\rightarrow 2$ kind of elastic channels,
for which we get $\eta/s\approx 0.25-3$ near the Mott temperature.
In present work, we have focussed on $1\leftrightarrow 2$ kind of in-elastic channels
of quarks and mesons, which are calculated from the imaginary part of their
self-energies at finite temperature. For quark self-energy, quark-pion and quark-sigma meson
loops are taken while pion-sigma and pion-pion loops are taken to obtain pion and
sigma meson self-energies respectively.
The thermal widths of quark and mesons are
found to be non-zero in limited regions of high and low temperature domain because of
their respective threshold conditions. Therefore, when we
estimate different transport coefficients using those thermal widths, we are
able to predict their values within those temperature ranges only.
Interestingly, we notice that near (but above)
the Mott temperature, $\eta/s$ can reach the KSS value ($\sim 0.08$)
because of such $1\rightarrow 2$ inelastic scattering processes. The contribution of inelastic channels
to the calculation of thermal widths may be comparable, or may even be dominant, to those of $2 \rightarrow 2$ elastic scatterings,. Although beyond a certain temperature,
the contribution of in-elastic scattering vanishes and only elastic
channels~\cite{HM_PQM2} survive leding to low $\eta/s$.
%


The present work of two flavor PQM model can be generalized to a more realistic 2 +1 flavor model. With more number
of mesons there are many more channels involving the strange quarks. It will be interesting to look at the different branch cut
structure due to the difference in  masses of the strange and non strange quarks. Calculations in this regard are in progress and will be reported in future.

{\bf Acknowledgment:} PS is supported from the scheme of DST Inspire (India).
SG acknowledges to IIT-Bhilai, funded by Ministry of Human
Resource Development (MHRD), Govt. of India.
SG would like to thank to Sandeep Chatterjee, Bhaswar Chatterjee, Snigdha Ghosh and Purushottam Sahu for
getting some help on this work. Authors are very thankful to Kinkar Saha
and Rajarshi Ray for their useful comments. SG, AA and HM acknowledge WHEPP-2017 for getting
some fruitful discussions. GK is financially
supported by the DST-INSPIRE faculty award under Grant No.
DST/INSPIRE/04/2017/002293.

\end{document}